\documentclass[a4paper,11pt]{article}
\pdfoutput=1 

\usepackage{jheppub} 

\usepackage[T1]{fontenc} 

\title{\boldmath Domain Wall Constraints on Two Higgs Doublet Models with $Z_2$ Symmetry}

\author{Richard A. Battye}
\author{Apostolos Pilaftsis}
\author{and Dominic G. Viatic}
\affiliation{Department of Physics and Astronomy, University of Manchester, Manchester M13 9PL, United\\Kingdom}

\emailAdd{richard.battye@manchester.ac.uk}
\emailAdd{apostolos.pilaftsis@manchester.ac.uk}
\emailAdd{dominic.viatic@postgrad.manchester.ac.uk}

\abstract{The Two Higgs Doublet Model (2HDM) with spontaneously broken $Z_2$ sym\-metry predicts a production of domain walls at the electroweak scale.
We derive cosmo\-logical constraints on model parameters 
for both Type-I and Type-II 2HDMs from the requirement that domain walls
do not dominate the Universe by the present day. For Type-I 2HDMs, we 
deduce the lower bound on the key parameter $\tan\beta > 10^5$ for a wide range of Higgs-boson masses $\sim$ 100 GeV or greater close to the Standard Model alignment limit.
In addition, we perform numerical simulations of the 2HDM with an approximate
as well as an exact $Z_2$ symmetry but biased initial conditions.
In both cases, we find that domain wall networks are unstable and, hence, do not survive at late times.
The domain walls experience an exponential suppression of scaling in these models which can help ameliorate the stringent constraints found in the case of an exact discrete symmetry.
For a 2HDM with softly-broken $Z_2$ symmetry, we relate the size of this exponential suppression to the soft-breaking bilinear parameter~$m_{12}$ 
allowing limits to be placed on this parameter of order $\mu$eV,
such that domain wall domination can be avoided.
In particular, for Type-II 2HDMs, we obtain a corresponding lower limit on the CP-odd phase $\theta$ generated by QCD instantons, $\theta \ \stackrel{>}{{}_\sim}\ 10^{-11}/(\sin\beta \cos\beta)$, which is in some tension with the upper limit of $\theta \ \stackrel{<}{{}_\sim}\ 10^{-11}$--$10^{-10}$,
as derived from the non-observation of a non-zero neutron electric dipole moment.
For a $Z_2$-symmetric 2HDM with biased initial conditions, we are able to relate the size of the exponential suppression to a biasing parameter $\varepsilon$ so as to avoid domain wall domination.
}

\begin{document} 
\maketitle
\flushbottom

\section{Introduction}
\label{sec:intro}

Domain walls are topological defects which emerge from the breaking of discrete sym\-metries~\cite{Garagounis2003} resulting in a vacuum manifold containing topologically disconnected points.
These disconnected points correspond to degenerate vacua.
Phase transitions producing domain walls occur at a finite rate and as such, the field can select different vacua in causally disconnected regions of space.
This divides the universe into so-called \textit{domains} the interfaces between which are \textit{domain walls} \cite{Shellard1994}.
Defects which emerge from the breaking of a global symmetry are expected to enter a regime of dynamical scaling such that the number of defects is constant per Hubble horizon \cite{Nakayama2017}.
Domain walls follow a power law scaling with an exponent close to 1 as shown in simulations for the so-called Goldstone model with a single real scalar field \cite{Garagounis2003,Pearson2009}.
This is due to the fact that the walls have a tension under which they collapse as quickly as causality permits.
This scaling feature results in an undesirable fate for the Universe where domain walls can be present in nature, in stark contrast to all our  observations.
The energy density of matter and radiation both scale proportionally to $(\text{time})^{-2}$ in their respective epochs of domination.
However, domain wall energy density scales proportionally to $(\text{time})^{-1}$ \cite{Lazanu2015}.
This means that domain walls will come to dominate the Universe at late times \cite{Zeldovich1974,Larsson1996,Press1989}.
This is the so-called \textit{domain wall problem}.
Therefore, if cosmic domain walls are to exist in nature constraints must be placed on domain wall-forming models such that domain wall domination does not occur \cite{Lazanu2015} or, at least, occurs after present day.
Domination could be avoided by, for example, having an additional field couple to the walls altering their scaling behaviour \cite{Pearson2009,Pearson2010,Viatic2020}.
Alternatively, one could have the domain walls decay before they come to dominate the Universe by making the discrete symmetry approximate.
It is important to note that these modifications require a change in the symmetries of the model and must, therefore, be well-motivated given the fundamental role of symmetries in physics.

The Higgs mechanism for electroweak symmetry breaking was verified by the measurement of a Higgs boson \cite{Aad2012,Chatrchyan2012} of mass 125 GeV at the LHC \cite{Aad2015}.
The properties of this scalar particle so far match those predicted for the SM Higgs scalar~\cite{Djouadi:2005gi,Khachatryan2016}.
Nonetheless, current experimental measurements do not prohibit the existence of more scalar particles.
One minimal and theoretically well-motivated extension which can be made to the SM is to introduce a second complex Higgs doublet into the theory.
This is the so-called \textit{Two Higgs Doublet Model} (2HDM) first suggested by T. D. Lee in 1973 \cite{Lee1973}. A complete study of the general 
CP-violating 2HDM is given in~\cite{Pilaftsis1999} (for a review, see \cite{Branco2012}).

It is well known that the 2HDM predicts the emergence of a variety of topological defects, such as domain walls, vortices and global monopoles, from the breaking of accidental symmetries which the model can posses under certain parameter choices \cite{Brawn2011,Eto:2018hhg,Eto:2018tnk,Chen:2020soj}.
In the thermal history of the Universe, theories of New Physics, such as the 2HDM,  can predict a series of symmetry breaking phase transitions as the Universe expanded and cooled \cite{Shellard1994}.
These broken symmetries are no longer observable but should be restored in the early Universe when temperatures were far higher than at present~\cite{Garagounis2003}.
These phase transitions can leave relic topological defects which can serve as probes of high energy physics in the early Universe \cite{Garagounis2003,Kibble:1982dd,Nakayama2017}.

In this article we focus our attention on the 2HDM with $Z_2$ symmetry, whose spontaneous breaking predicts the existence of domain wall solutions.
In particular, we obtain cosmo\-logical constraints on the theoretical  parameters of the 2HDM which arise from the non-observation of such domain 
walls. By solving the relevant equations of motion${}$~\cite{Garagounis2003,Moss2006,Pearson2009,Pearson2010},
we present a number of  numerical simulations of such topological defects that consolidate the cosmological constraints derived in this paper. 

The remainder of this article is structured as follows.
In Section \ref{sec:2HDM}, we outline the scalar sector of the 2HDM with softly-broken $Z_2$ symmetry, introduce the physical degrees of freedom in the model and provide a brief review of experimental constraints on the 2HDM for models of Type I and II.
In Section \ref{sec:DWconstraints}, we present constraints on the 2HDM from domain wall domination for cases where the $Z_2$ symmetry is exact and parameter regimes in which domain wall domination could be avoided.
In Sections \ref{sec:ApproxSymm} and \ref{sec:BiasedICs}, we present results of simulations of the 2HDM with both an approximate $Z_2$ symmetry and biased initial conditions for a 2HDM where the $Z_2$ symmetry is exact.
In both cases, domain wall networks are unstable and domain wall domination can be avoided by requiring that these domain wall networks be sufficiently short-lived.
Finally, Section \ref{sec:discussion} summarizes the main results of our study.

\newpage

\section{\boldmath The Two Higgs Doublet Model with Softly-Broken $Z_2$ Symmetry}
\label{sec:2HDM}

Under a $Z_2$ transformation the complex scalar Higgs doublets, $\Phi_1$ and $\Phi_2$, transform as
\begin{equation}
    \Phi_1 \rightarrow \Phi_1,\quad \Phi_2 \rightarrow -\Phi_2.
\end{equation}
The 2HDM potential with softly-broken $Z_2$ symmetry can be written as
\begin{equation}\label{eq:brokenZ2Potential}
    \begin{split}
    V = & -\mu_1^2 \Phi_1^\dagger\Phi_1 - \mu_2^2 \Phi_2^\dagger\Phi_2 - m_{12}^2(\Phi_1^\dagger\Phi_2 + \Phi_2^\dagger\Phi_1) + \lambda_1 (\Phi_1^\dagger\Phi_1)^2 + \lambda_2 (\Phi_2^\dagger\Phi_2)^2 \\
    & + \lambda_3 (\Phi_1^\dagger\Phi_1)(\Phi_2^\dagger\Phi_2) + \lambda_4(\Phi_1^\dagger\Phi_2)(\Phi_2^\dagger\Phi_1) - \frac{\left|\lambda_5\right|}{2}\left[(\Phi_1^\dagger\Phi_2)^2 + (\Phi_2^\dagger\Phi_1)^2\right]
    \end{split}
\end{equation}
with 8 real parameters: $\mu_{1}^2, \mu_2^2, m_{12}^2, \lambda_1, \lambda_2, \lambda_3, \lambda_{4}$ and $\lambda_5$.
The field bilinear $\Phi_1^\dagger\Phi_2$ violates the $Z_2$ symmetry and hence this model possesses an approximate $Z_2$ symmetry for small values of the coefficient $m_{12}^2$.
Moreover, in the limit $m_{12}^2 = 0$ \eqref{eq:brokenZ2Potential} possesses an \textit{exact} $Z_2$ symmetry.

The vacua are parametrized as
\begin{equation}\label{eq:NormalVac}
	\langle\Phi_1\rangle = \frac{1}{\sqrt{2}}\left(\begin{matrix}
	0 \\
	v_1
	\end{matrix}\right),\quad 
	\langle\Phi_2\rangle = \frac{1}{\sqrt{2}}\left(\begin{matrix}
	0 \\
	v_2
	\end{matrix}\right).
\end{equation}
This parametrization will be referred to as \textit{CP-preserving vacua}.
The parameters $ v_1,v_2,v_+ $ and $ \xi $ are referred to as the \textit{vacuum manifold parameters}.
For the CP-preserving vacua \eqref{eq:NormalVac}, the VEVs can be calculated in terms of the potential parameters,
\begin{equation}\label{eq:Z2VEVs}
    v_1^2 = \frac{4\lambda_2\mu_1^2 - 2\tilde{\lambda}_{345}\mu_2^2}{4\lambda_1\lambda_2 - \tilde{\lambda}_{345}^2},\quad
    v_2^2 = \frac{4\lambda_1\mu_2^2 - 2\tilde{\lambda}_{345}\mu_1^2}{4\lambda_1\lambda_2 - \tilde{\lambda}_{345}^2},
\end{equation}
where we have defined $\tilde{\lambda}_{345} = \lambda_3 + \lambda_4 - \left|\lambda_5\right|$.

The 2HDM has 5 physical scalar particles: 2 neutral CP-even states, $h$ and $H$, one CP-odd neutral state, $A$, and 2 charged states, $H^\pm$.
The other three scalar degrees of freedom correspond to would-be Goldstone bosons, $G^0$ and $G^\pm$, which are absorbed into the longitudinal components of of the electroweak gauge bosons, $W^\pm$ and $Z^0$.
We identify $h$ with the Higgs particle measured at the LHC by ATLAS and CMS \cite{Aad2012,Chatrchyan2012} fixing the parameter $M_h$ at 125 GeV \cite{Aad2015}.
Furthermore, the SM VEV is fixed at $v_{\text{SM}} = $ 246 GeV.

In order to investigate the evolution of domain walls in the 2HDM with approximate $Z_2$ symmetry we first obtain a physical parametrization of the model with which to perform our numerical simulations.
Expressions for the masses of the scalar Higgs particles, $h$, $H$, $A$ and $H^\pm$, are obtain as eigenvalues of the Hessian matrix of \eqref{eq:brokenZ2Potential} using the parametrization
\begin{equation}\label{eq:HiggsDoublets}
    \Phi_1 = \left(\begin{matrix}
    \varphi_1^+ \\
    \frac{1}{\sqrt{2}}(v_1 + \varphi_1 + i a_1)
    \end{matrix}\right), \quad
    \Phi_2 = \left(\begin{matrix}
    \varphi_2^+ \\
    \frac{1}{\sqrt{2}}(v_2 + \varphi_2 + i a_2)
    \end{matrix}\right),
\end{equation}
where $\varphi_i^+$ are complex scalar fields.
The CP-even mass matrix is derived to be
\begin{equation}
    \mathcal{M}_N^2 = \left(\begin{matrix}
    m_{12}^2 \tan\beta + 2 \lambda_1 c_\beta^2 v_\text{SM}^2 & -m_{12}^2 + \tilde{\lambda}_{345} v_\text{SM}^2 s_\beta c_\beta \\
    -m_{12}^2 + \tilde{\lambda}_{345} v_\text{SM}^2 s_\beta c_\beta & m_{12}^2\cot\beta + 2 \lambda_2 s_\beta^2 v_\text{SM}^2
    \end{matrix}\right),
\end{equation}
while the CP-odd mass is
\begin{equation}
    M_A^2 = \frac{m_{12}^2}{s_\beta c_\beta} + \left|\lambda_5\right| v_\text{SM}^2
\end{equation}
where we have introduced the short-hand notations $\sin x = s_x$ and $\cos x = c_x$.
The charged Higgs mass can then be written as
\begin{equation}
    M_{H^\pm}^2 = M_A^2 - \frac{1}{2}(\lambda_4 + \left|\lambda_5\right|) v_\text{SM}^2.
\end{equation}
The CP-even mass matrix is diagonalized by the mixing angle, $\alpha$,
\begin{equation}
    \mathcal{M}_N^2 = \left(\begin{matrix}
    c_\alpha & -s_\alpha \\
    s_\alpha & c_\alpha
    \end{matrix}\right)
    \left(\begin{matrix}
    M_h^2 & 0 \\
    0 & M_H^2
    \end{matrix}\right)
    \left(\begin{matrix}
    c_\alpha & s_\alpha \\
    -s_\alpha & c_\alpha
    \end{matrix}\right).
\end{equation}
Using the above expressions one obtains the physical parametrization of the scalar potential
\begin{align} \label{eq:softZ2params}
    \mu_1^2 &= -m_{12}^2 \tan\beta + \frac{1}{2}(M_h^2 c_\alpha^2 + M_H^2 s_\alpha^2) + \frac{1}{2}(M_h^2 - M_H^2) c_\alpha s_\alpha \tan\beta,\nonumber\\
    \mu_2^2 &= -m_{12}^2 \cot\beta + \frac{1}{2}(M_h^2 s_\alpha^2 + M_H^2 c_\alpha^2) + \frac{1}{2}(M_h^2 - M_H^2) c_\alpha s_\alpha \cot\beta,\nonumber\\
    \lambda_1 &= \frac{-m_{12}^2 \tan\beta + M_h^2 c_\alpha^2 + M_H^2 s_\alpha^2}{2 c_\beta^2 v_\text{SM}^2},\nonumber\\
    \lambda_2 &= \frac{-m_{12}^2 \cot\beta + M_h^2 s_\alpha^2 + M_H^2 c_\alpha^2}{2 s_\beta^2 v_\text{SM}^2},\\
    \lambda_3 &= \frac{-m_{12}^2 + 2M_{H^\pm}^2 c_\beta s_\beta + (M_h^2 - M_H^2) c_\alpha s_\alpha}{c_\beta s_\beta v_\text{SM}^2},\nonumber\\
    \lambda_4 &= \frac{m_{12}^2 + (M_A^2 - 2M_{H^\pm}^2)c_\beta s_\beta}{c_\beta s_\beta v_\text{SM}^2},\nonumber\\
    \left|\lambda_5\right| &= \frac{-m_{12}^2 + M_A^2 c_\beta s_\beta}{c_\beta s_\beta v_\text{SM}^2}\nonumber.
\end{align}
Rescaling for dimensionless energy per unit area, $\hat{E} = E/M_h v_\text{SM}^2$, we can write \eqref{eq:brokenZ2Potential} in a dimensionless form,
\begin{equation}
    \begin{split}
    \Hat{V} = &  - \frac{1}{2}\left[-2\hat{m}^2 \tan\beta + c_\alpha^2 + \Hat{M}_H^2 s_\alpha^2 + (1 - \Hat{M}_H^2) c_\alpha s_\alpha \tan\beta \right] \Hat{\Phi}_1^\dagger \Hat{\Phi}_1 \\
    & - \frac{1}{2}\left[-2\hat{m}^2 \cot\beta + s_\alpha^2 + \Hat{M}_H^2 c_\alpha^2 + (1 - \Hat{M}_H^2) c_\alpha s_\alpha \cot\beta \right] \Hat{\Phi}_2^\dagger \Hat{\Phi}_2 - \hat{m}^2(\hat{\Phi}_1^\dagger\hat{\Phi}_2 + \hat{\Phi}_2^\dagger\hat{\Phi}_1) \\
    & + \frac{-\hat{m}^2 \tan\beta + c_\alpha^2 + \Hat{M}_H^2 s_\alpha^2}{2 c_\beta^2} (\Hat{\Phi}_1^\dagger \Hat{\Phi}_1)^2
    + \frac{-\hat{m}^2 \cot\beta + s_\alpha^2 + \Hat{M}_H^2 c_\alpha^2}{2 s_\beta^2} (\Hat{\Phi}_2^\dagger \Hat{\Phi}_2)^2 \\
    & + \frac{(1 - \Hat{M}_H^2) c_\alpha s_\alpha -\hat{m}^2 + 2\hat{M}_{H^\pm}^2 c_\beta s_\beta}{c_\beta s_\beta} (\Hat{\Phi}_1^\dagger \Hat{\Phi}_1)(\Hat{\Phi}_2^\dagger \Hat{\Phi}_2) \\
    & + \frac{\hat{m}^2 + (\hat{M}_A^2 - 2\hat{M}_{H^\pm}^2)c_\beta s_\beta}{c_\beta s_\beta} (\Hat{\Phi}_1^\dagger \Hat{\Phi}_2)(\Hat{\Phi}_2^\dagger \Hat{\Phi}_1) + \frac{\hat{m}^2 - \hat{M}_A^2 c_\beta s_\beta}{2 c_\beta s_\beta} \left[(\Hat{\Phi}_1^\dagger \Hat{\Phi}_2)^2 + (\Hat{\Phi}_2^\dagger \Hat{\Phi}_1)^2\right],
    \end{split}
\end{equation}
with 6 dimensionless parameters $\hat{M}_H,\hat{M}_A,\hat{M}_{H^\pm},\alpha,\beta$ and $\hat{m}^2$. The dimensionless masses are scaled by the SM Higgs mass, i.e. $\hat{M}_i \equiv M_i/M_h$ and $\hat{m}^2 \equiv m_{12}^2/M_h^2$, and $\alpha/\beta$ are the CP-even/odd mixing angles which diagonalize the scalar mass matrices.
For details of this reparametrization and rescaling procedure, see \cite{Viatic2020}.

In order to have a phenomenologically acceptable model, we must also consider that current measurements of signal rates for the Higgs discovered at the LHC are close to those predicted by the SM \cite{Djouadi:2005gi,Khachatryan2016}.
Therefore, we must restrict our investigation to parameters in/near the so-called \textit{SM alignment limit} where the couplings of the CP-even scalar, $h$, match those predicted by the SM.
Exact SM alignment is obtained when the relation $\cos\left(\alpha - \beta\right) = 1$ holds\footnote{Constraints from experimental limits on the mixing angles for varying values of the Higgs masses can be found in \cite{Arbey2018,Haller2018,Chowdhury2017,Han2017}.}.
Therefore, the physical parametrization we will use in phenomenological discussions to follow will be $\left\{M_h,M_H,M_A,M_{H^\pm},v_\text{SM}^2,\tan\beta,\cos\left(\alpha-\beta\right)\right\}$.

When choosing physical parameters and considering the implications of our phenomenology in later sections we must also account for the effect of model type on current experimental constraints on the 2HDM.
The \textit{Type} of a 2HDM is determined by the form of its Yukawa sector.
In other words, constraints on the masses of the 2HDM scalars from experimentally-measured signal rates are type-dependent due to differences in the Higgs-fermion interactions of the models.
The Yukawa Lagrangian can be written in its most general form as
\begin{equation}\label{eq:Yukawa2HDM}
    \mathcal{L}_{\text{Y}} = -\sum_{i=1}^{2} \left(\bar{q}_L i\sigma^2 \Phi_i Y_i^u u_R + \bar{q}_L \Phi_i Y_i^d d_R + \bar{\ell}_L\Phi_i Y_i^\ell e_R + \text{h.c.}\right),
\end{equation}
where $q_L$ and $\ell_L$ are SU(2)$_L$ doublets for left-handed quarks and leptons, repsectively; $u_R$, $d_R$ and $\ell_R$ are SU(2)$_L$ singlets for right-handed up-type quarks, down-type quarks and leptons, respectively\footnote{Note that all these objects are 3-vectors is flavour space where flavour indices have been suppressed.} and $Y_i^{u,d,\ell}$ are $3\times3$ Yukawa matrices for each Higgs doublet.
However, the 2HDM Yukawa sector as given in \eqref{eq:Yukawa2HDM} is too general and restrictions on the Higgs-fermion couplings are required to remove or limit tree-level flavour-changing neutral currents (FCNC) \cite{Branco2012}.
The two models we will consider are the so-called \textit{Type I} and \textit{Type II} 2HDMs \cite{Ivanov2017}.
In Type I models all fermions couple to only one of the doublets (conventionally chosen to be $\Phi_2$) \cite{Branco2012} whereas in Type II models, up-type quarks couple to one doublet ($\Phi_2$ by convention) and down-type quarks and leptons to the other \cite{Arbey2018} \footnote{The specific forms of the Higgs-fermion couplings in each case can be found in Table. 2 of \cite{Branco2012} in terms of the mixing angles, $\alpha$ and $\beta$.}.

For a 2HDM with softly broken $Z_2$ symmetry of Type I 2HDM, flavour physics constraints place a limit of $\tan\beta>1$ for $M_{H\pm} = 1$ TeV with the constraint strengthening to $\tan\beta>3$ for $M_{H^\pm} = 100$ GeV \cite{Arbey2018,Haller2018}.
As such, the entire range of masses from $100$ GeV upwards can be chosen from $M_{H^\pm}$ without contradicting flavour constraints provided sufficiently large values for $\tan\beta$ are chosen.
Constraints on the charged Higgs from direct and indirect detection at the LHC increase the lower bound on $\tan\beta$ for $M_{H^\pm} < 300$ GeV \cite{Arbey2018}.
For a Type II 2HDM flavour physics constraints place much stronger bounds on the charged Higgs mass.
Specifically, a lower bound of $M_{H^\pm} \gtrsim 600$ GeV exists for all $\tan\beta$ and $M_{H^\pm} \gtrsim 650$ GeV for $\tan\beta < 1$ independent of the other physical 2HDM parameters \cite{Arbey2018}. 
In both Type I and Type II 2HDMs the combined constraints of \cite{Haller2018} require a strong alignment between $M_{H^\pm}$ and either $M_H$ or $M_A$.
This is attributed to the strong constraint on the value of $\cos(\alpha-\beta)$ from current Higgs coupling measurements being close to SM alignment.
As with the charged Higgs constraints, the Type II model is more strongly constrained by current observations with the entire parameter range of the neutral Higgs masses, $100 \; \text{GeV} \le M_{H,A} \le 1000 \; \text{GeV}$ considered in \cite{Haller2018}, ruled out for their lower benchmark charged Higgs masses of $M_{H^\pm}=250$ GeV and $500$ GeV.
Therefore, we choose alignment of the charged Higgs mass with either the scalar mass, $M_H$, or the pseudoscalar mass, $M_A$, motivated by the combined constraints of \cite{Haller2018}.

\section{\boldmath Exact $Z_2$ Symmetry}
\label{sec:DWconstraints}

In an FRW universe, the energy density of both matter and radiation decrease proportionally to $(\text{time})^{-2}$ in their respective epochs of domination.
However, domain wall energy density decreases proportionally to $(\text{time})^{-1}$.
Therefore, the energy density of domain walls will increase relative to matter and radiation in their respective epochs and, hence, come to dominate the energy density of the universe.
The time of this domination is determined by the energy per unit area of the domain walls.
It is clear that we do not live in a domain wall dominated universe and, therefore, any model which produces domain walls must not allow domination before present day.
It has been established that domain wall networks in the $Z_2$-symmetric 2HDM exhibit a deviation from $\propto t^{-1}$ scaling \cite{Viatic2020}.
Specifically, more domain walls are predicted at late times in the 2HDM than one would expect for $\propto t^{-1}$ scaling.
This feature makes the domain wall problem more restrictive.
Here, we calculate the domain wall density for the $Z_2$-symmetric 2HDM assuming $\propto t^{-1}$ scaling and require that domain wall domination occurs after present day to obtain corresponding constraints on the physical observables.
As such, this calculation provides a minimal constraint from 2HDM domain wall domination.

In the $Z_2$-symmetric 2HDM, the energy per unit area of the domain walls is given by
\begin{equation}
    E = \int_{-\infty}^\infty \mathcal{E}(x) dx,
\end{equation}
where
\begin{equation}\label{eq:Z2enden}
    \mathcal{E}(x) = \frac{1}{2}\left(\frac{d v_1}{dx}\right)^2 + \frac{1}{2}\left(\frac{d v_2}{dx}\right)^2 - \frac{1}{2} \mu_1^2 v_1^2 - \frac{1}{2}\mu_2^2 v_2^2 + \frac{1}{4}\lambda_1 v_1^4 + \frac{1}{4}\lambda_2 v_2^4 + \frac{1}{4} \tilde{\lambda}_{345} v_1^2 v_2^2
\end{equation}
for CP-preserving vacua \eqref{eq:NormalVac}.
The topologically non-trivial solution which minimizes the energy per unit area can be obtained via gradient flow (see, for example \cite{Brawn2011,Viatic2020}).
It should be noted that the SM VEV, $v_\text{SM} = \sqrt{v_1^2 + v_2^2}$, also changes in the vicinity of the kink as the solution interpolates from one vacuum to another.
The energy density of a domain wall network can be approximated by a self-scaling argument.
The total energy within a Hubble horizon of radius, $r$ is proportional to $E r^2$.
Therefore, the energy density, $\rho_\text{dw} \propto E r^{-1}$ and since the horizon expands at the speed of light, $\rho_\text{dw} \propto E t^{-1}$.
Hence, we write
\begin{equation}
    \rho_\text{dw} = A \hat{E} M_h v_\text{SM}^2 t^{-1}
\end{equation}
where $A$ is a constant of proportionality, quantifying the number of walls per horizon,
and define a corresponding domain wall density parameter in the usual manner,
\begin{equation}
    \Omega_{\text{dw}} = \frac{\rho_{\text{dw}}}{\rho_{\text{crit}}},
\end{equation}
with critical density at present day
\begin{equation}
    \rho_\text{crit}(t_0) = \frac{3 H_0^2 M_\text{pl}^2}{8 \pi},
\end{equation}
where we have used $H_0 = 72 \text{km} \ \text{s}^{-1} \ \text{Mpc}^{-1} = 1.54 \times 10^{-42} \ \text{GeV}$ in natural units.
For $\Omega_\text{dw} < 1$ at present day, we obtain the limit
\begin{equation}
    \frac{8\pi A \hat{E} M_h v_\text{SM}^2}{3 H_0^2 t_0 M_\text{pl}^2} < 1.
\end{equation}
Therefore, for $t_0 \simeq 6.6 \times 10^{41} \ \text{GeV}^{-1}$ and $M_\text{pl} \simeq 1.2 \times 10^{19}$ GeV, we obtain the dimensionless inequality,
\begin{equation}\label{eq:exactIneq}
    A \hat{E} < \frac{3 H_0^2 t_0 M_\text{pl}^2}{8 \pi M_h v_\text{SM}^2} \simeq 3.6 \times 10^{-12}.
\end{equation}

\begin{figure}
    \centering
    \includegraphics[width = \textwidth]{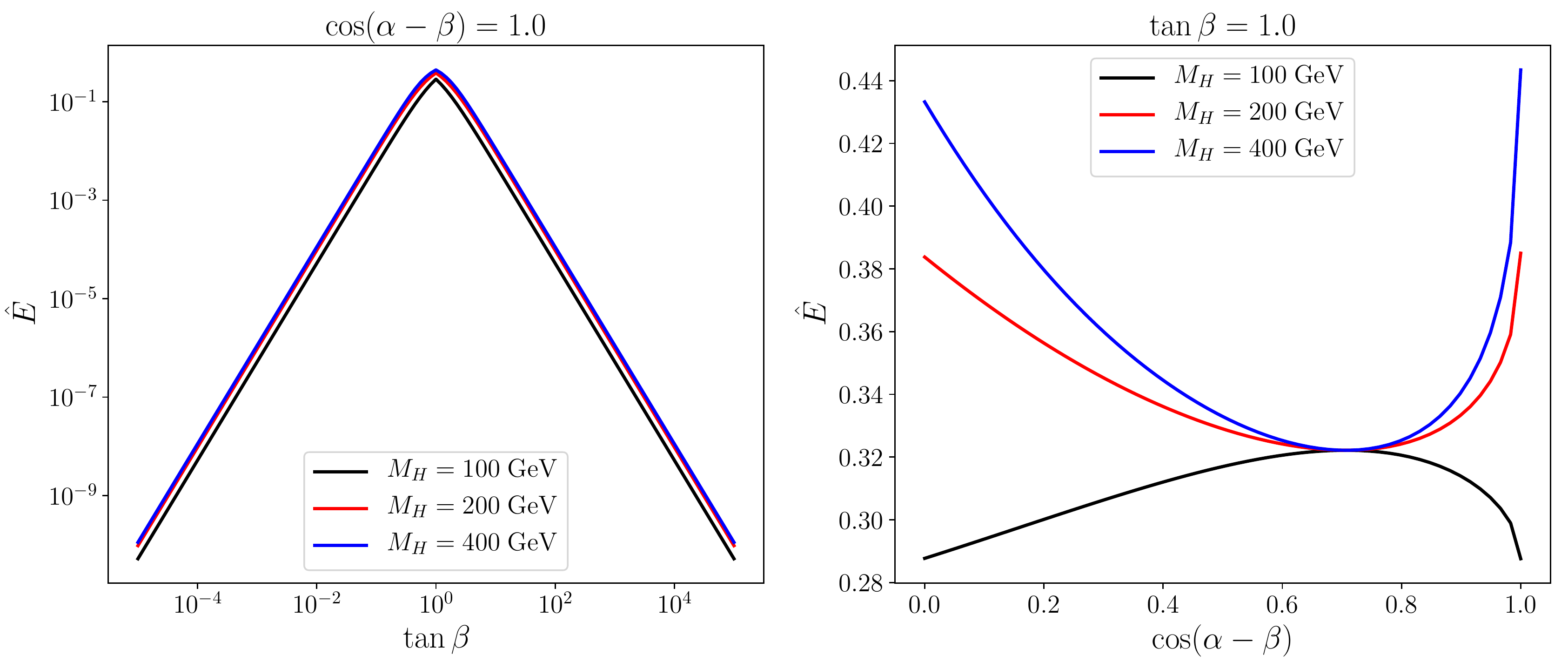}
    \caption{Variation of dimensionless energy per unit area of minimum energy kink solutions with $\tan\beta$ (left) and the SM alignment parameter, $\cos(\alpha - \beta)$ (right), for benchmark values of the CP-even scalar mass, $M_H$. In all cases the energy per unit area of kink solutions goes to zero in the limit of large $\tan\beta$.}
    \label{fig:Etanb}
\end{figure}

Firstly, it should be noted that agreement with this limit can always be obtained for sufficiently large or small values of $\tan\beta$.
It is always energetically favourable for the kink to interpolate between the smaller of the two VEVs.
Since $\tan\beta$ determines the relative size of the VEVs, for $\tan\beta > 1$ the first doublet is $Z_2$ odd while for $\tan\beta <1$ the second doublet is $Z_2$ odd.
This is illustrated in Fig.~\ref{fig:Etanb} for some benchmark values of the CP-even scalar mass in the SM alignment limit.
We see in the left panel of Fig.~\ref{fig:Etanb} that domain walls do indeed become \textit{ultra-light} in large and small limits of $\tan\beta$ where the VEV of the $Z_2$ odd doublet becomes vanishingly small.
The parameter $\tan\beta$ is the primary parameter in the variation of $\hat{E}$.
In the right-hand panel of Fig.~\ref{fig:Etanb} we see that the alignment parameter $\cos(\alpha-\beta)$ only has a weak effect on the energy density.
Moreover, the impact of changing $M_H$ is also weak.

\begin{figure}
    \centering
    \includegraphics[width = \textwidth]{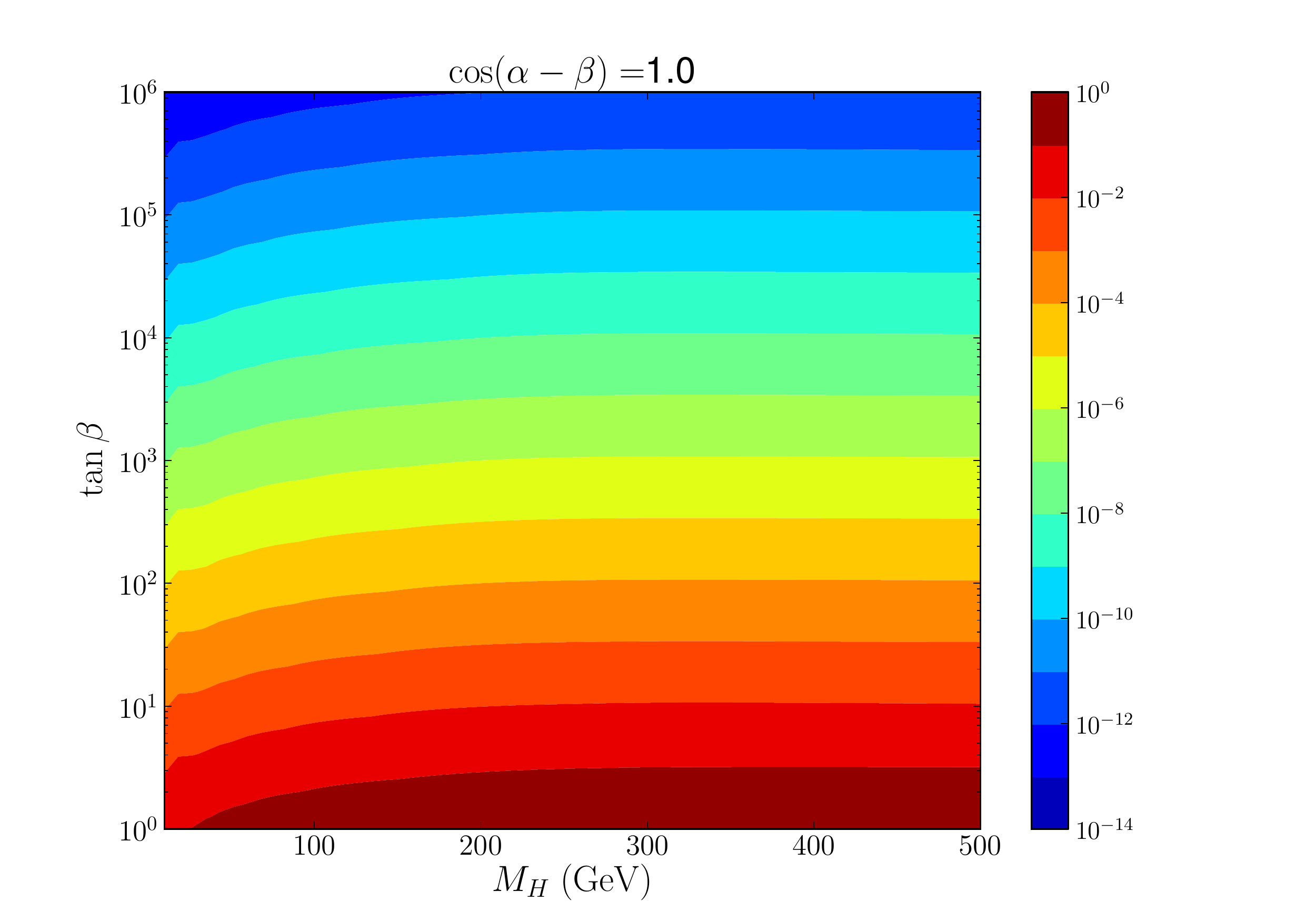}
    \caption{Variation of the dimensionless energy per unit area in the $Z_2$-symmetric 2HDM with SM alignment limit as a function of the CP-even scalar mass, $M_H$, and CP-odd mixing angle, $\tan\beta = v_2/v_1$, for minimum energy kink solutions obtained via gradient flow. Contour regions indicate the order of magnitude of the energy per unit area for a given set of physical parameters.}
    \label{fig:envar}
\end{figure}

The variation of dimensionless energy per unit area, $\hat{E}$ in SM alignment is given in Fig.~\ref{fig:envar}.
Note that there is some subtlety in the impact of the parameter $A$ on this limit.
This is more than simply an assumed number of domain walls per horizon as this value will change between the matter and radiation eras through which these domain walls scale.
Nonetheless, this should note affect the order of magnitude of the limit \eqref{eq:exactIneq}.
We find that for domain walls of the type seen in the neutral vacuum minimum energy kink solutions a domain wall problem can only be avoided for experimentally viable Higgs masses at large values of $\tan\beta$ (of the order $10^5$ or more).
In lower $\tan\beta$ regimes one cannot evade the constraints placed on the $Z_2$-symmetric 2HDM by domain wall domination without unreasonably low values of the scalar masses.
It should be made clear that these results only pertain to scenarios where the 2HDM possess an \textit{exact} discrete symmetry and, hence, a domain wall problem.
These stringent constraints suggest that in order to have cosmologically viable 2HDM domain walls in experimentally viable parameter regimes a means of modifying the scaling behaviour of these domain walls will be required.

\section{\boldmath Approximate $Z_2$ Symmetry}
\label{sec:ApproxSymm}

So far, we have only considered the scenario in which the 2HDM possesses an \textit{exact} $Z_2$ symmetry.
We have shown that the domain wall problem present in this model places highly restrictive limits on the parameters of the model such that domination can be made to occur after present day.
We now turn our attention to means of eliminating the domain wall problem altogether.
For the 2HDM with softly-broken $Z_2$ symmetry, \eqref{eq:brokenZ2Potential}, the degeneracy of the vacua is removed and the scalar potential contains so-called true and false vacua.
The true vacuum is the global minimum of the potential while the false vacuum is a local minimum with higher energy.
We anticipate that the energy difference between these vacua produces a pressure on the domains of false vacuum causing the domain walls to collapse when this pressure becomes comparable to the surface tension of the walls \cite{Larsson1996}.
Therefore, the domain wall problem could be eliminated in this scenario if domain wall networks are sufficiently short-lived that they do not survive long enough to dominate the energy density of the universe.
We earlier made the self-scaling argument that domain wall energy density can be expressed as $\rho_\text{dw} \propto E t^{-1}$.
Let us add an exponential suppression to this domain wall energy we have ${\rho}_\text{dw} \propto E t^{-1} e^{-\alpha t}$ with corresponding density parameter,
\begin{equation}\label{eq:softdensity}
    \Omega_{\text{dw}} \propto \frac{E t}{M_\text{Pl}^2} e^{-\alpha t},
\end{equation}
where the parameter $\alpha$ encodes the breaking of the symmetry and we will estimate this in the 2HDM.
Again, introducing a dimensionless proportionality constant, $A$, specifying the number of domain walls per horizon and recalling that in our dimensionless system the energy per unit area, $E = M_h v_\text{SM}^2 \hat{E}$ where $M_h =$ 125 GeV and $v_\text{SM} =$ 246 GeV, we can write
\begin{equation}\label{eq:dwdens}
    \Omega_{\text{dw}} = \frac{32\pi}{3}\frac{A \hat{E} M_h v_\text{SM}^2}{M_\text{Pl}^2} t e^{-\alpha t}.
\end{equation}
The time at which the exponential suppression dominates is determined by the parameter $\alpha$ after which time the domain wall density relative to critical begins to decrease.
In other words, the domain wall density is maximal at $t_\text{max} = 1/\alpha$.
Therefore, the maximum value is given by\footnote{To avoid confusion note that, here, $e$ is Euler's constant \textit{not} the fundamental electric charge.}
\begin{equation}
    \Omega_\text{dw}^\text{max} \equiv \Omega_\text{dw}(t_\text{max}) = \frac{32\pi}{3}\frac{A\hat{E} M_h}{e \alpha} \left(\frac{v_\text{SM}}{M_\text{Pl}}\right)^2.
\end{equation}
The minimal theoretical requirement is that $\Omega_\text{dw}^\text{max} < 1$ such that the domain walls do not dominate the energy density of the universe by the end of their scaling phase.
Inserting numerical values into \eqref{eq:dwdens} we obtain the lower bound
\begin{equation}\label{eq:alphalimit}
    \alpha > \alpha_\text{min} \simeq \frac{64\pi}{3} A \hat{E} \times 10^{-32} \; \text{GeV}
\end{equation}
from which we find the time of maximum domain wall density,
\begin{equation}
    t_\text{max} = \frac{3}{64\pi}\frac{10^{32} \ {\text{GeV}}^{-1}}{A \hat{E}} = \frac{9.8\times10^{5} \ \text{secs}}{A \hat{E}}.
\end{equation}
It has been established that kink solutions in the $Z_2$-symmetric 2HDM have dimensionless energy of order 1 for physically viable values of physical observables \cite{Viatic2020}.
Note that the time of radiation-matter equality, $t_\text{eq} \sim 10^{12}$ seconds.
As such, it is apparent that even the smallest acceptable exponential suppression, $\alpha_\text{min}$, places the time of maximum domain wall density well within the radiation dominated epoch provided domain walls are not ultra-light, i.e. for large $\tan\beta$ where $\hat{E}$ becomes small.
It should also be noted that the collapse of these domain walls could still have undesirable effects on the Cosmic Microwave Background (CMB) and Big Bang Nucleosynthesis (BBN) \cite{Chen:2020soj} conflicting with current cosmological constraints on these processes.
One may wish to consider constraints on domain walls arising at these epochs, however, the domain wall density \eqref{eq:softdensity} suggests that domain wall domination in the 2HDM would not have arisen by the BBN epoch.
Moreover, without choosing particular combinations of physical parameters, e.g. ultra-light domain walls, such defects will have already come to dominate the universe prior to recombination.

\begin{figure}
    \centering
    \hspace*{-0.71cm}
    \includegraphics[width=\textwidth]{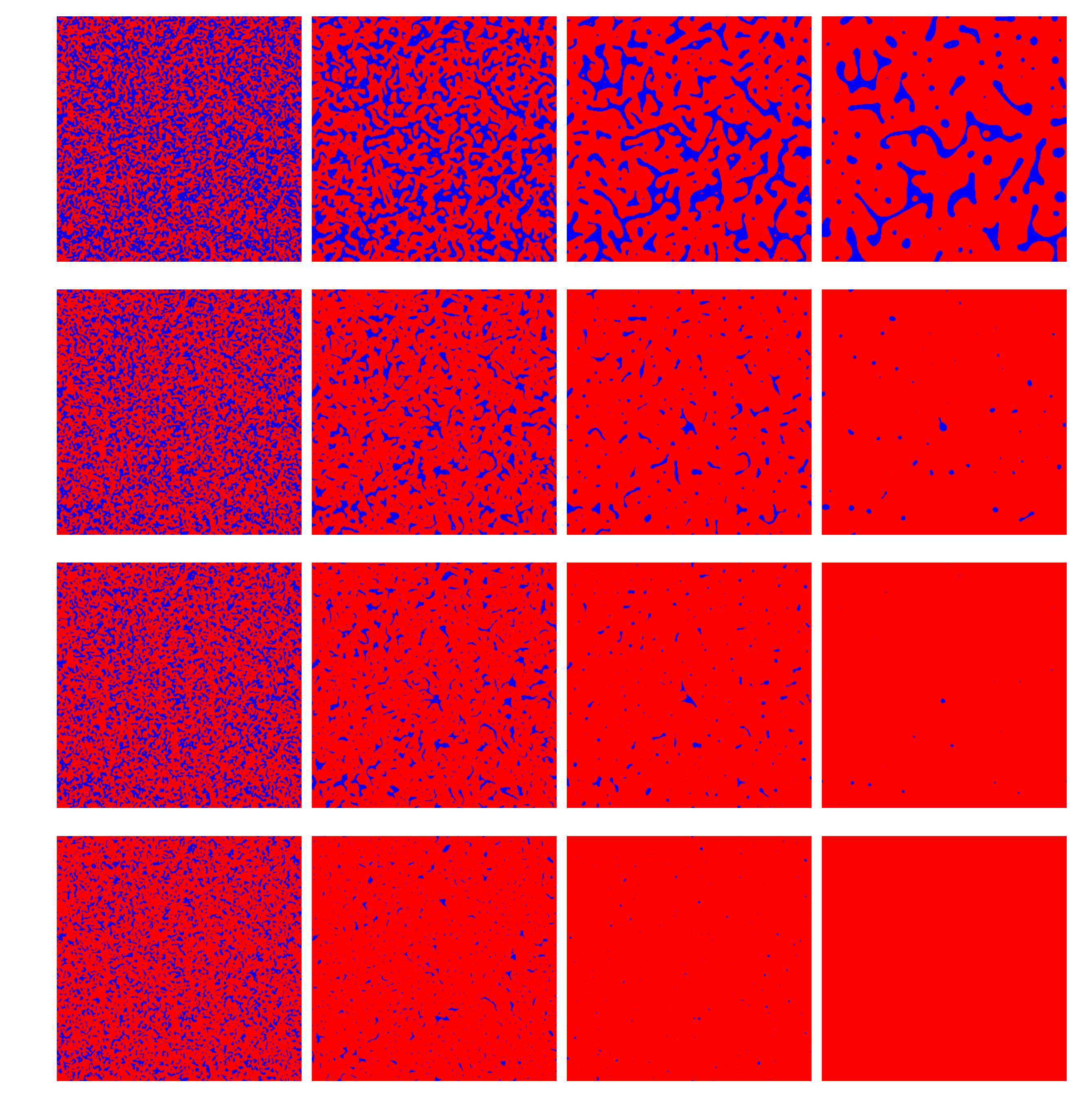}
    \caption{Evolution of domain walls in a 2HDM with approximate $Z_2$ symmetry in (2+1) dimensions for dimensionless soft-breaking parameter, $\hat{m}^2 = 2 \times {10}^{-3}$, $4 \times {10}^{-3}$, $5 \times {10}^{-3}$ and $7 \times {10}^{-3}$ top-to-bottom. Remaining parameters are common to all sets of simulations and were chosen as $M_H = M_A = M_{H^\pm} = 200\;\text{GeV}$, $\tan\beta = 0.85$, $\cos(\alpha - \beta) = 1.0$.
    Simulations were run for time, $t=448$ with temporal grid spacing, $\Delta t = 0.2$ and spatial grid size, $P = 4096$ with spacing, $\Delta x = 0.9$.
    Each set of plots progress in time left-to-right and each plot is at double the timestep of the previous.}
    \label{fig:softZ2domains}
\end{figure}

\begin{figure}
    \centering
    \includegraphics[width=\textwidth]{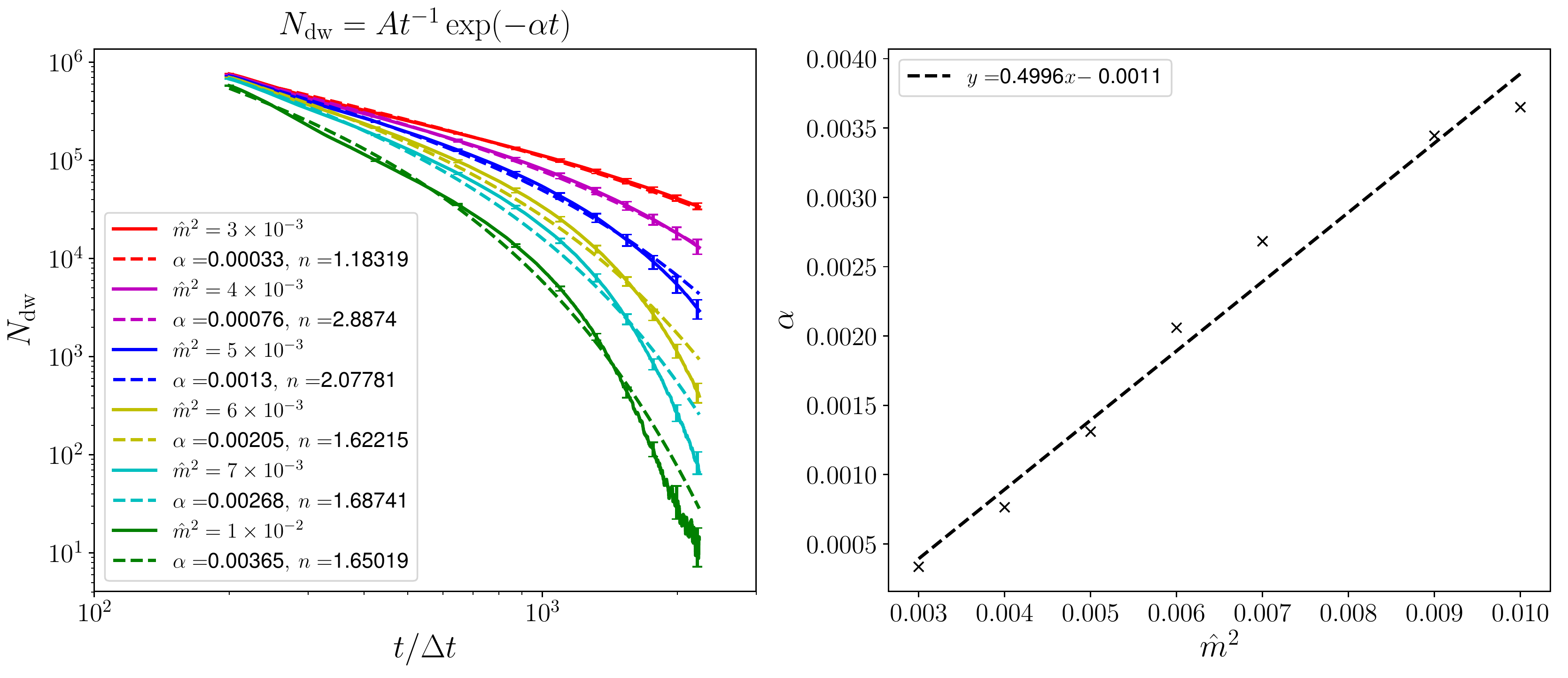} \\
    \includegraphics[width=\textwidth]{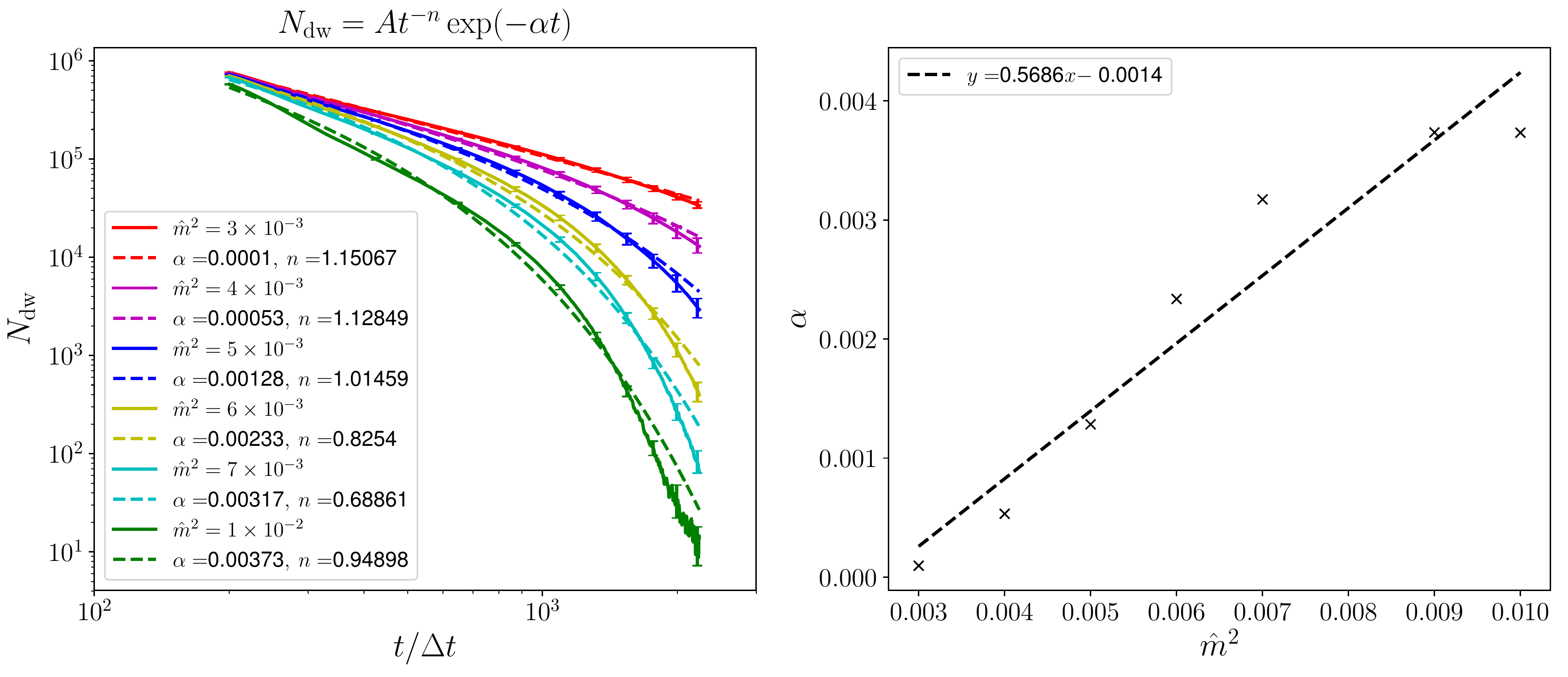}
    \caption{Evolution of the number of domain walls in 2D 2HDM simulations with approximate $Z_2$ symmetry averaged over 10 realizations for various values of the dimensionless soft breaking parameter, $\hat{m}^2$.
    Remaining parameters chosen were $M_H = M_A = M_{H^\pm} = 200\;\text{GeV}$, $\tan\beta = 0.85$ and $\cos(\alpha - \beta) = 1.0$.
    Simulations were run for time, $t=1260$ with temporal grid spacing, $\Delta t = 0.2$ and spatial grid size, $P = 4096$ with spacing, $\Delta x = 0.9$.
    Error bars, which are the standard deviation amongst the realizations, illustrate the numerical scatter.
    Also shown are non-linear least squares fittings for an exponentially suppressed power law for $n=1$ fixed (top) and allowing $n$ to vary (bottom).}
    \label{fig:brokenZ2scaling}
\end{figure}

We anticipate that limits on the suppression coefficient, $\alpha$, will allow constraints on the soft-breaking parameter $\hat{m}^2$ to be obtained such that the domain walls can be made cosmologically benign.
We have performed (2+1) dimensional simulations for the global scalar field theory of the 2HDM with approximate $Z_2$ symmetry on a regular grid of $P^2$ points for $P=4096$ with Minkowski metric (for details of the simulation procedure see \cite{Viatic2020}).
Simulations are performed in (2+1) dimensions for computational ease.
Nonetheless, these simulations are a good approximation of the behaviour in (3+1) dimensions, as shown in \cite{Viatic2020}.
The evolution of a set of such simulation is presented in Fig.~\ref{fig:softZ2domains} for various values of the soft breaking parameter, $\hat{m}^2$.
In these simulations domain walls are short-lived with the entire space coming to be dominated by a single vacuum at late times.
The time at which the field comes to occupy the true vacuum throughout the space is determined by the size of the symmetry breaking term, $\hat{m}^2$.
This collapse has a significant effect on the scaling behaviour of domain walls in the approximately $Z_2$ symmetric 2HDM.
The number of domain walls as a function of time in (2+1) dimensions are presented in Fig.~\ref{fig:brokenZ2scaling} for various values of the dimensionless soft-breaking parameter, $\hat{m}^2$.
The time evolution of the number of domain walls is obtained as an average over 10 realizations.
Fig.~\ref{fig:brokenZ2scaling} shows that the number of domain walls in the 2HDM with approximate $Z_2$ symmetry decreases much more rapidly than the $t^{-1}$ scaling found in models with exact discrete symmetries \cite{Garagounis2003}.
The number of domain walls appears to follow an exponentially suppressed power law, $N_\text{dw} \propto t^{-n} e^{-\alpha t}$.
A non-linear least squares fitting to the number of domain walls for an exponentially suppressed power law are also included in Fig.~\ref{fig:brokenZ2scaling} for both $n=1$ and $n$ as a fitted parameter.
The exponential suppression parameter, $\alpha$, exhibits the approximate relationship to the soft-breaking parameter,
\begin{equation}
    \frac{\alpha}{M_h} \simeq 0.5 \ \hat{m}^2 = 0.5 \left(\frac{m_{12}}{M_h}\right)^2.
\end{equation}
Hence, in order for the exponential suppression of domain wall density to be sufficiently large to avoid domination, we obtain the limit
\begin{equation}
   \label{eq:m2limit}
    m_{12}^2 > \frac{64 \ \pi}{3}\frac{A \hat{E}}{e} \left(\frac{v_\text{SM} M_h}{M_\text{Pl}}\right)^2 \simeq 1.6\times10^{-28} A \hat{E} \ \text{GeV}^2.
\end{equation}
Assuming a small number of domain walls per horizon such that $A$ is of order unity, this limit suggests a small value of $m_{12}$ relative to the electroweak scale (around $\mu\text{eV}$ order) would sufficiently modify domain wall scaling to avoid their domination.

In a $Z_2$-symmetric Type-II 2HDM, the origin of a small effective $m^2_{12}$ parameter may be attributed to anomalous QCD instanton effects~\cite{tHooft:1976rip,tHooft:1976snw}. In the absence of a Peccei--Quinn mechanism~\cite{Peccei:1977hh,Peccei:1977ur},
we may nonetheless adapt their results and
conservatively estimate the size of the anomalous $Z_2$-breaking parameter $m^2_{12}$ from the would-be PQ instanton potential as follows:
\begin{equation}
\begin{split}
   \label{eq:Vinst}
V_{\rm inst} &\sim \Lambda^4_{\rm QCD}\,\Bigg[
\bigg(\frac{\Phi^\dagger_1 \Phi_2}{v^2_{\rm SM}}\bigg)^{n_G}\: -\: 
\bigg(\frac{\Phi^\dagger_1 \Phi_2\, e^{i\theta}}{v^2_{\rm SM}}\bigg)^{n_G}\Bigg]\ +\ {\rm H.c.} \\
&\stackrel{<}{{}_\sim} \frac{\Lambda^4_{\rm QCD}}{v^2_{\rm SM}}\, s^2_\beta c^2_\beta\, \Big(1 - \cos\big(n_G\theta\big)\Big)\; \Phi^+_1\Phi_2\ +\ {\rm H.c.}\,, 
\end{split}
\end{equation}
where $\Lambda_{\rm QCD} \sim 0.3$~GeV is the QCD confinement scale, 
$n_G=3$ is the number of the SM quark generations,
and $\theta$ is the well-known strong CP-odd phase
generated by QCD instantons. A non-zero value 
of $\theta$ would induce a non-zero Electric Dipole Moment (EDM) for the neutron~\cite{Kim:2008hd}.
Current experiments place an upper limit on $\theta \stackrel{<}{{}_\sim} 10^{-10}$--$10^{-11}$~\cite{PhysRevLett.124.081803}.
On the other hand, combining~\eqref{eq:Vinst}
with~\eqref{eq:m2limit}, we obtain a lower limit on $\theta$,
\begin{equation}
  \label{eq:theta}  
\theta \ \stackrel{>}{{}_\sim}\ \frac{10^{-11}}{s_\beta c_\beta} \; .   
\end{equation}
This suggests that the parameter $\tan\beta$ should lie in the narrow interval: $0.3 \stackrel{<}{{}_\sim} \tan\beta \stackrel{<}{{}_\sim} 3$.

\section{Biased Initial Conditions}
\label{sec:BiasedICs}

One can also avoid domain wall domination in models with an exact discrete symmetry by biasing the initial conditions such that the degenerate vacua are selected with unequal probability.
In many studies of domain wall dynamics, including our own simulations of 2HDM domain walls \cite{Viatic2020}, it is assumed that domain walls evolve from initial conditions where each of the degenerate vacua are selected with equal probability, i.e. that the scalar field(s) are in thermal equilibrium before the phase transition \cite{Coulson1995}.
However, if the initial conditions in the early Universe have some bias towards one of the vacua, smaller domains of the disfavoured vacuum should form surrounded by larger regions where the field lies in the preferred vacuum \cite{Larsson1996}.
These small domain walls should then collapse rapidly.
This has been demonstrated for the Goldstone model in (2+1) and (3+1) dimensions \cite{Coulson1995}.
Of course, these initial conditions must be viable for the Higgs fields in the early Universe if such biased 2HDM domain walls are to be of interest for cosmology.

\begin{figure}
    \centering
    \hspace*{-0.71cm}
    \includegraphics[width=\textwidth]{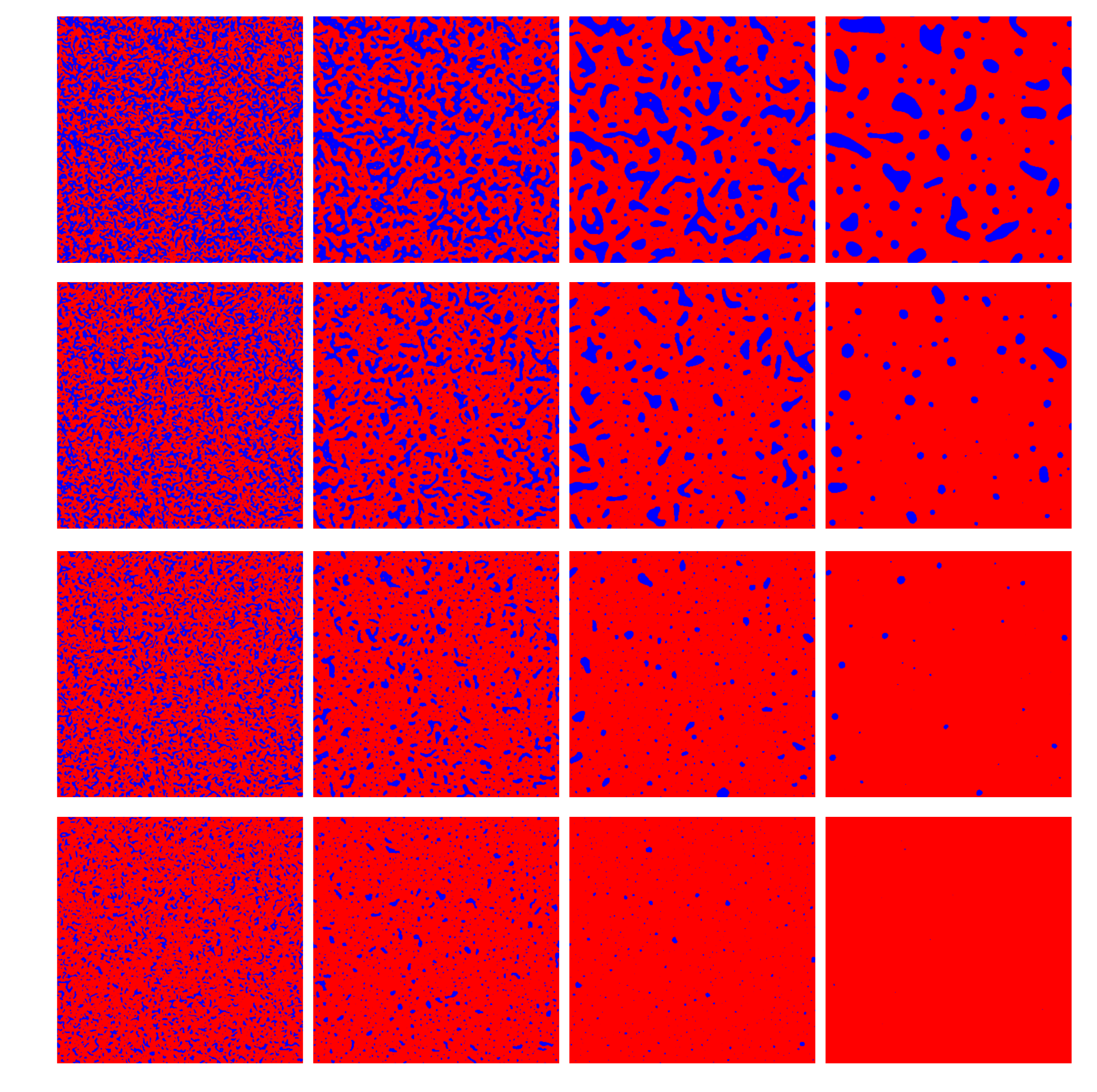}
    \caption{2D simulations of the evolution of domain walls in a 2HDM with $Z_2$ symmetry from increasingly biased initial conditions top-to-bottom. Parameters chosen were $M_H = M_A = M_{H^\pm} = 200\;\text{GeV}$, $\tan\beta = 0.85$ and $\cos(\alpha - \beta) = 1.0$.
    Simulation was run for time, $t=448$ with temporal grid spacing, $\Delta t = 0.2$ and spatial grid size, $P = 4096$ with spacing, $\Delta x = 0.9$.
    Each set of plots progress in time left-to-right and each plot is at double the timestep of the previous.}
    \label{fig:biasedZ2domains}
\end{figure}

\begin{figure}
    \centering
    \includegraphics[width=\textwidth]{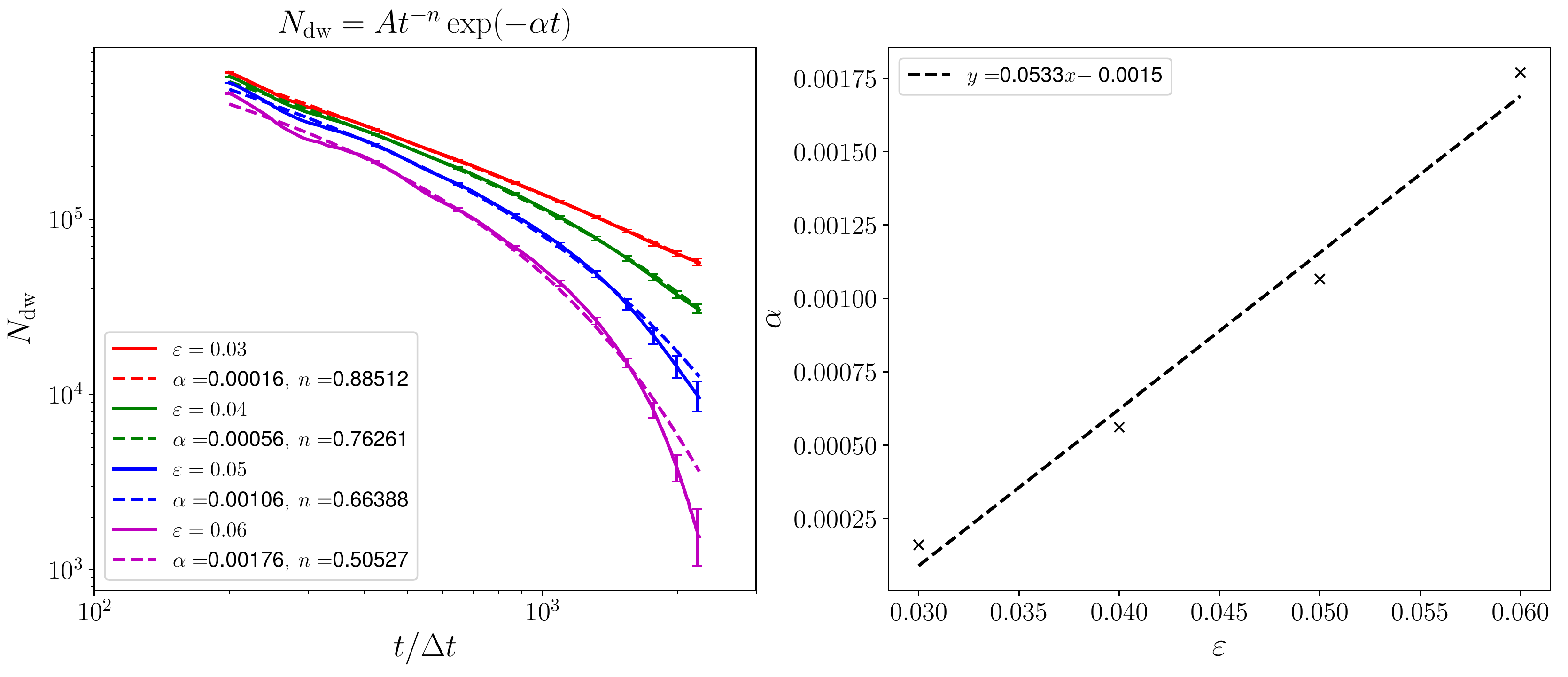}
    \caption{Evolution of the number of domain walls in 2D 2HDM simulations with $Z_2$ symmetry from biased initial conditions (i.e. normally distributed about $\varepsilon$) averaged over 10 realizations.
    Also plotted is the standard power law scaling for a domain wall network $\propto t^{-1}$.
    Parameters chosen were $M_H = M_A = M_{H^\pm} = 200\;\text{GeV}$, $\tan\beta = 0.85$ and $\cos(\alpha - \beta) = 1.0$.
    Simulations were run for time, $t=448$ with temporal grid spacing, $\Delta t = 0.2$ and spatial grid size, $P = 4096$ with spacing, $\Delta x = 0.9$.
    Error bars, which are the standard deviation amongst the realizations, illustrate the numerical scatter.}
    \label{fig:biasedZ2scaling}
\end{figure}

Assuming an exponential suppression of domain wall scaling, the lower bound of \eqref{eq:alphalimit} still holds in this case.
The aim now becomes to relate this to the bias parameter, $\varepsilon$.
We have performed (2+1) dimensional simulations with $P=4096$ for the global scalar field theory of the 2HDM with exact $Z_2$ symmetry with Minkowski metric from biased random initial conditions.
Specifically, we produce random initial conditions for the scalar fields normally distributed around $\varepsilon$ such that one vacuum is selected with greater probability.
The evolution of a set of such simulation is presented in Fig.~\ref{fig:biasedZ2domains}.
We find that domain wall are short-lived with the entire space coming to be dominated by the preferred vacuum at late times.
This behaviour is qualitatively similar to that found in the softly-broken $Z_2$ case of Fig.~\ref{fig:softZ2domains}.
The number of domain walls as a function of time in (2+1) dimensions for biased initial conditions are presented in Fig.~\ref{fig:biasedZ2scaling}.
The time evolution of the number of domain walls is obtained as an average over 10 realizations.
Fig.~\ref{fig:biasedZ2scaling} shows the number of domain walls decreasing with a similar profile to that seen in the case of approximate symmetry.
The number of domain walls appears to follow an exponentially suppressed power law, $N_\text{dw} \propto t^{-n} e^{-\alpha t}$.
A non-linear least squares fitting to the number of domain walls for an exponentially suppressed power law are also included in Fig.~\ref{fig:biasedZ2scaling}.
The exponential suppression parameter, $\alpha$, shows an approximate linear relationship to the biasing,
\begin{equation}
    \frac{\alpha}{M_h} \simeq 0.05 \frac{\varepsilon}{v_\text{SM}}
\end{equation}
Hence, in order for the exponential suppression of domain wall density to be sufficiently large to avoid domination, we obtain the limit
\begin{equation}
  \label{eq:varepsilon}
    \varepsilon > \frac{640\pi}{3} \frac{A \hat{E}}{e}\left(\frac{v_\text{SM}^{3/2}}{M_\text{Pl}}\right)^2 \simeq 2.5 \times 10^{-29} A \hat{E} \ \text{GeV}.
\end{equation}
Again, assuming a small number of domain walls per horizon such that $A$ is of order unity, this limit suggests a very small biasing of the initial conditions would be sufficient to avoid domain wall domination.


\section{Conclusions}
\label{sec:discussion}

In this article we have considered the phenomenological implications of domain walls in 2HDMs with an exact or approximate $Z_2$ symmetry.
We have obtained cosmological constraints on the Higgs masses and mixing angles such that domain walls in these models do not dominate the energy density of the Universe today.
We find that domain wall domination can always be avoided for sufficiently large or small values of $\tan\beta$ where domain walls become \textit{ultra-light}, i.e. the energy of the neutral vacuum solution tends to zero.
Moreover, for Type-I 2HDMs with spontaneous breakdown of the $Z_2$ symmetry, we find that domain wall domination can only be avoided today for $\tan\beta > 10^5$ for scalar masses larger than~100~GeV. 

We have also demonstrated that domain wall networks in (2+1) dimensional simulations can be made to collapse by rendering the discrete symmetry approximate via a small symmetry breaking term.
We find that the time evolution of the number of such domain\- walls exhibits an exponential suppression of the approximate power law scaling found in~\cite{Viatic2020}.
The collapse rate of the domain walls is linearly related to the soft-breaking parameter squared,~$m^2_{12}$.
Consequently, we find that a soft-breaking parameter $m_{12} \sim 10^{-6}\,$eV is sufficient to avoid domain wall domination by the end the scaling phase of their evolution.
For a 2HDM of Type II, this suggests a corresponding lower limit on the CP-odd phase~$\theta$ generated by QCD instantons, i.e.~$\theta \ \stackrel{>}{{}_\sim}\ 10^{-11} / s_\beta c_\beta$. This estimate is
in some tension with an upper limit on $\theta \stackrel{<}{{}_\sim} 10^{-10}$--$10^{-11}$ coming from the non-observation of a non-zero
EDM for the neutron. Taking this last constraint into account, we obtain
an upper and lower limit on the key parameter $\tan\beta$, i.e.~$0.3 \stackrel{<}{{}_\sim} \tan\beta \stackrel{<}{{}_\sim} 3$.
We anticipate similar exponential suppression of domain wall scaling will be obtained in 2HDMs with approximate CP1 and CP2 symmetries, or in any alternative scenario which breaks the $Z_2$ symmetry.

Finally, we have demonstrated that domain walls evolving from biased initial conditions can be similarly short-lived with qualitatively similar behaviour to the case of an approximate symmetry.
We find that domain walls in our biased simulations also experiencing an exponential suppression of their scaling.
In particular, we have derived in~\eqref{eq:varepsilon} a lower limit on the biasing parameter~$\varepsilon$ of initial conditions such that domain wall domination can be avoided by the end of scaling.
Results obtained for approximate and biased discrete symmetries can both provide means of avoiding the late-time scaling problems which domain walls ordinarily present, and hence they could be used to make 2HDMs with discrete symmetries cosmologically safe.

\bigskip

 \appendix

\acknowledgments

The work of RB and AP are supported in part by the Lancaster-Manchester-Sheffield Consortium for Fundamental Physics under STFC research grant ST/L000520/1.


\bibliographystyle{JHEP}
\bibliography{bibliography}



\end{document}